\documentclass[11pt,reqno]{amsart}
\usepackage{amsfonts}
\usepackage{mathrsfs}
\usepackage{amsmath}
\usepackage{amssymb,amsfonts}
\usepackage{cite}

\newtheorem{thm}{Theorem}[section]
\newtheorem{lem}{Lemma}[section]

\newtheorem{prop}{Proposition}[section]
\numberwithin{equation}{section}
\newtheorem{rmk}{Remark}[section]

\def\pf{{\textit {Proof:} }}

\newcommand{\mysection}[1]{\section{#1}\setcounter{equation}{0}}

\newfont{\bb}{msbm10 at 11pt}

\newcommand{\bal}{\begin{aligned}}      \newcommand{\eal}{\end{aligned}}
\newcommand{\ba}{\begin{array}}      \newcommand{\ea}{\end{array}}
\newcommand{\bc}{\begin{center}}     \newcommand{\ec}{\end{center}}
\newcommand{\be}{\begin{enumerate}}  \newcommand{\ee}{\end{enumerate}}
\newcommand{\beq}{\begin{eqnarray}}  \newcommand{\eeq}{\end{eqnarray}}
\newcommand{\beQ}{\begin{eqnarray*}} \newcommand{\eeQ}{\end{eqnarray*}}
\newcommand{\bi}{\begin{itemize}}    \newcommand{\ei}{\end{itemize}}
\newcommand{\bt}{\begin{tabular}}    \newcommand{\et}{\end{tabular}}
\newcommand{\bdm}{\begin{displaymath}} \newcommand{\edm}{\end{displaymath}}

\def\qed{\hfill{Q.E.D.}\smallskip}

\newcommand{\ls}{\setlength{\baselineskip}{12pt}
                 \setlength{\parskip}{3mm}}

\begin{document}

\allowdisplaybreaks

\title[Massless Majorana spinors in the Kerr Spacetime]{Massless Majorana spinors in the Kerr spacetime}

\author[T Cai]{Tianyuan Cai$^{\dag}$}
\address[]{$^{\dag}$School of Mathematics and Information Science, Guangxi University, Nanning, Guangxi 530004, PR China}
\email{caitianyuan@st.gxu.edu.cn}
\author[X Zhang]{Xiao Zhang$^{\flat}$}
\address[]{$^{\flat}$ State Key Laboratory of Mathematical Sciences, Academy of Mathematics and Systems Science, Chinese Academy of Sciences, Beijing 100190, PR China}
\address[]{$^{\flat}$ School of Mathematical Sciences, University of Chinese Academy of Sciences, Beijing 100049, PR China}
\address{$^{\flat}$Guangxi Center for Mathematical Research, Guangxi University, Nanning, Guangxi 530004, PR China}
\email{xzhang@amss.ac.cn}

\date{}

\begin{abstract}
In this paper, we show that massive Majorana spinors \eqref{1.2} do not exist if they are $t$-dependent or $\phi$-dependent in Kerr, or Kerr-(A)dS spacetimes. For massless Majorana spinors in the non-extreme Kerr spacetime, the Dirac equation can be separated into radial and angular equations, parameterized by two complex constants $\epsilon_1$, $\epsilon_2$. If at least one of $\epsilon_1$, $\epsilon_2$ is zero, massless Majorana spinors can be solved explicitly. If $\epsilon_1$, $\epsilon_2$ are nonzero, we prove the nonexistence of massless time-periodic Majorana spinors in the non-extreme Kerr spacetime which are $L^p$ outside the event horizon for $	0<p\le\frac{6}{|\epsilon_1|+|\epsilon_2| +2}$. We then provide the Hamiltonian formulation for massless Majorana spinors and prove that the self-adjointness of the Hamiltonian leads to the angular momentum $a=0$ and spacetime reduces to the Schwarzschild spacetime, moreover, the massless Majorana spinor must be $\phi$-independent. Finally, we show that, in the Schwarzschild spacetime, for initial data with $L^2$ decay at infinity, the probability of the massless Majorana spinors to be in any compact region of space tends to zero as time tends to infinity.
\end{abstract}

\maketitle \pagenumbering{arabic}

\mysection{Introduction}\ls

In general relativity, the Dirac equation is given by
\begin{equation}\label{1.1}
	D \Psi   +i\lambda\Psi=0
\end{equation}
for 4-dimensional Lorentzian manifolds, where $\Psi$ is a 4-dimensional complex spinor and $\lambda$ is a real number referred as the mass of the spinor $\Psi$. The spinor $\Psi$ is massive if $\lambda \neq 0$, and massless if $\lambda = 0$. 

In 1976, Chandrasekhar observed that, for the following time-periodic spinors in the Kerr spacetime 
\begin{align}\label{Dirac separation}
\Psi=S^{-1}\psi,\quad\quad \psi=e^{-i  \left(\omega t+(k+\frac{1}{2})\phi\right)} \left(\begin{array}{c}
R_{-}(r)\Theta_{-}(\theta)\\
R_{+}(r)\Theta_{+}(\theta)\\
R_{+}(r)\Theta_{-}(\theta)\\
R_{-}(r)\Theta_{+}(\theta)
\end{array}\right),
\end{align}
where $S$ is a diagonal matrix
\begin{align} 
	S=\Delta_{r} ^{\frac{1}{4}}
	\text{diag}\left((r-ia\cos\theta)^{\frac{1}{2}}I_{2\times2},\,\,\,\,(r+ia\cos\theta)^{\frac{1}{2}}I_{2\times2}\right),\label{S-matrix}
\end{align}
the Dirac equation can be separated into radial and angular equations \cite{SC}. Page extended this result to the Kerr-Newman spacetime \cite{PD}. By using Chandrasekhar's separation, Finster, Kamran, Smoller and Yau proved nonexistence of normalizable time-periodic solutions in the non-extreme Kerr-Newman spacetime \cite{FJ}. This indicates that the normalizable Dirac particles either disappear into the black hole or escape to infinity. Moreover, they studied the probability for Dirac particles to be inside a given annulus located outside the event horizon and proved that it tends to zero as $t\rightarrow \infty$ \cite{FK}. They also derived decay rates and probability estimates that massive Dirac particles escape to infinity \cite{FF}. Some related works can be found in \cite{BC09, BC10, WZ18, FWZ24, BS06, BS07, B07, FC}.

It is interesting whether these properties hold true for Majorana spinors, which were proposed by Majorana in 1937 \cite{EM} and studied extensively in order to search for neutrinos in the universe in recent years, eg \cite{GT, NJ, RW, AA, MS, PU, HM}. However, Chandrasekhar's separation of spinors is not consistent with the Majorana condition. In \cite{ZZ}, a new time-periodic spinor
\begin{equation}\label{1.2}
	\Psi=S^{-1}\psi,\quad\quad  \psi=E\begin{pmatrix}
		R_-(r)\Theta_-(\theta)\\
		R_+(r)\Theta_+(\theta)\\
		\overline{R}_+(r)\overline{\Theta}_+(\theta)\\
		-\overline{R}_-(r)\overline{\Theta}_-(\theta)
	\end{pmatrix},
\end{equation}
was proposed, where $S$ is given by \eqref{S-matrix} and 
\begin{align}
	E=\text{diag}\left(e^{-i(\omega t+(k+\frac{1}{2})\phi)}I_{2\times2},\,\,\,\, e^{i(\omega t+(k+\frac{1}{2})\phi)}I_{2\times2}\right). \label{E-matrix}
\end{align}
The Dirac equation \eqref{1.1} for Majorana spinor \eqref{1.2} can be reduced to four equations in Kerr-Newman and Kerr-Newman-(A)dS spacetimes. Moreover, the four equations yield two algebraic equations which indicate nonexistence of differentiable Majorana spinors in these spacetimes if either the coupled magnetic charge $P_e\neq 0$ or the coupled electric charge $Q_e \neq 0$ \cite{ZZ, ZZ1}. It is therefore worth asking what happens to Majorana spinors in Kerr type spacetimes. We show that massive Majorana spinors \eqref{1.2} do not exist if they are $t$-dependent or $\phi$-dependent in Kerr, or Kerr-(A)dS spacetimes. For massless Majorana spinors in the non-extreme Kerr spacetime, the Dirac equation can be separated into radial and angular equations, parameterized by two complex constants $\epsilon_1$, $\epsilon_2$. If at least one of $\epsilon_1$, $\epsilon_2$ is zero, massless Majorana spinors can be solved explicitly. If $\epsilon_1$, $\epsilon_2$ are nonzero, we prove
\begin{thm}\label{thm 1.1}
	In the non-extreme Kerr spacetime, for any massless time-periodic solutions \eqref{1.2} of the Dirac equation \eqref{1.1}, then $|R|$ is bounded at the event horizon $r=r_e$. Moreover, if the solution $\Psi$ is $L^p$ outside the event horizon for certain 
	\begin{align*}
		0<p\le\frac{6}{|\epsilon_1|+|\epsilon_2| +2},
	\end{align*}
	where $\epsilon_1$, $\epsilon_2$ are some nonzero complex constants, $\Psi$ must be zero. In particular, any normalizable such spinor, i.e. $p=2$, must be zero if $|\epsilon_1|+|\epsilon_2|\leq 1$.
\end{thm}
We then provide the Hamiltonian formulation for massless Majorana spinors and prove that the self-adjointness of the Hamiltonian leads to both the angular momentum $a=0$ and spacetime reduces to the Schwarzschild spacetime, moreover, the massless Majorana spinor must be $\phi$-independent. Finally, we provide the decay of the probability in the Schwarzschild spacetime.
\begin{thm}\label{thm 1.2}
	Consider Cauchy problem for the Dirac equation in the Schwarzschild spacetime 
	\begin{equation*}
		D\Psi=0,\quad\Psi(0,x)=\Psi_0, 
	\end{equation*}
	with the initial data $\Psi_0 \in L^2\big((2m,\infty)\times S^2, dvol\big)$, where $dvol$ is the volume form on the hypersurface $t=\text{const}$. For any $\delta>0$, $\mu>2m+\delta$, the probability of massless Majorana spinors with $\epsilon_1\neq 0$, $\epsilon_2\neq 0$, $|\epsilon_1|+|\epsilon_2|>1$ to be inside the annulus 
	\begin{align*}
		K_{\delta,\mu}=\Big\{2m+\delta\le r\le \mu \Big\}
	\end{align*}
	tends to zero as time tends to infinity.
\end{thm}

The paper is organized as follows. In Section 2, we review the results in \cite{ZZ} and prove a new result that massive Majorana spinors do not exist if they are $t$-dependent or $\phi$-dependent in Kerr, or Kerr-(A)dS spacetimes. In Section 3, we show the nonexistence of massless Majorana spinors in the non-extreme Kerr spacetime which are $L^p$ for certain $p>0$ outside the event horizon. In Section 4, we transform the Dirac equation into the Hamiltonian form and provide several boundary conditions to make the Hamiltonian operator Hermitian on a radially finite interval. In Section 5, we show that the angular solution can be solved explicitly for Majorana spinors on a radially finite interval and provide the asymptotic analysis of the radial solution at the event horizon and infinity. In Section 6, we construct an integral representation for the propagator and conclude that the probability of massless Majorana spinors to be in any compact region of space tends to zero as $t$ tends to infinity.

\mysection{Nonexistence in the Kerr-Newman type spacetimes}\ls

In this section, we first review the main result proved in \cite{ZZ} that there are no differentiable time-periodic Majorana spinors in the Kerr-Newman and Kerr-Newman-(A)dS spacetimes if either coupled magnetic charge or electric charge is nonzero. (See also \cite{ZZ1} for some physical interpretation as well as corrections of some typos on signs of elements in matrix $S$ and in equations (13)-(16), indices of frame $e_1$, $e_2$, $e_3$.) Then we prove a new result that there are no differentiable time-periodic massive Majorana spinors in the Kerr and Kerr-(A)dS spacetimes.

In Boyer-Lindquist coordinates, the Kerr-Newman and Kerr-Newman-(A)dS metrics take the form
\begin{align*}
	ds^2=&-\frac{\Delta_r}{U}\left(dt-\frac{a\sin^2\theta}{\Xi}d\phi\right)^2+\frac{U}{\Delta_r}dr^2+\frac{U}{\Delta_\theta}d\theta^2\\
	                   &+\frac{\Delta_\theta\sin^2\theta}{U}\left(adt-\frac{r^2+a^2}{\Xi}d\phi \right)^2,
\end{align*}
where $\Lambda=-3\kappa^2$ is the cosmological constant, $\kappa$ is taken to be zero, real or pure imaginary respectively, and
\begin{align*}
	\Delta_r= & \left(r^2+a^2\right)\left(1+\kappa^2 r^2\right)-2mr+P_e^2+Q_e^2,\\
	\Delta_\theta= & 1-\kappa^2a^2\cos^2\theta,\quad U= r^2+a^2\cos^2\theta,\quad \Xi=1-\kappa^2a^2>0.
\end{align*}
The metrics solve the Einstein-Maxwell field equations for the electromagnetic field
\begin{equation*}
	F_e=dA_e
\end{equation*}
where $A_e$ is the coupled electromagnetic potential
\begin{align*}
     A_e=-\frac{Q_er}{U}\left(dt-\frac{a\sin^2\theta}{\Xi}d\phi\right)-\frac{P_e\cos\theta}{U}\left(adt-\frac{r^2+a^2}{\Xi}d\phi\right),
\end{align*}
$Q_e$ is the coupled electric charge and $P_e$ is the coupled magnetic charge defined by 
\begin{equation*}
	Q_e=\frac{1}{4\pi}\oint_{S^2} \star F_e,\quad\quad P_e=\frac{1}{4\pi}\oint_{S^2}  F_e,
\end{equation*}
with the Hodge dual $\star F_e$ of $F_e$.

In the region $\Delta_r >0$, the coframe is chosen as
	\begin{align*}
			e^0&=\sqrt{\frac{\Delta_r}{U}}\left(dt-\frac{a\sin^2{\theta}}{\Xi}d{\phi}\right), \ \ \ \ e^1=\sqrt{\frac{U}{\Delta_\theta}}d{\theta}, \\
			e^2 &=\sqrt\frac{\Delta_\theta}{U}\sin{\theta} \left(a dt-\frac{r^2+a^2}{\Xi}d{\phi}\right),\ \ \ \
			e^3 =\sqrt{\frac{U}{\Delta_r}}d{r},
	\end{align*}
and the dual frame is chosen as
\begin{align*}
		e_0&=\frac{r^2+a^2}{\sqrt{U\Delta_r}}\left(\partial_t+\frac{a\Xi}{r^2+a^2}\partial_\phi\right), \ \ \ \
		e_1=\sqrt\frac{\Delta_\theta}{U}\partial_\theta,\\
		e_2&=-\sqrt\frac{1}{U\Delta_\theta}\left(a\sin\theta\partial_t+\frac{\Xi}{\sin\theta}\partial_\phi\right),\ \ \ \
		e_3=\sqrt{\frac{\Delta_r}{U}}\partial_r.
\end{align*}
The Cartan's structure equations are
\begin{align*}
		de^0&=C_{10}^0e^1\wedge e^0+C_{30}^0e^3\wedge e^0+C_{12}^0e^1\wedge e^2, \\
		de^1&=C_{31}^1e^3\wedge e^1,\\
		de^2&=C_{30}^2e^3\wedge e^0+C_{32}^2e^3\wedge e^2+C_{12}^2e^1\wedge e^2,\\
		de^3&=C_{31}^3e^3\wedge e^1,
\end{align*}
which give the connection 1-forms
\begin{align*}
		\omega^0_{\ 1}&=C_{10}^0e^0+\frac{1}{2}C_{12}^0e^2=-\omega_{01}, \ \
		\omega^0_{\ 2}=-\frac{1}{2}C_{30}^2e^3-\frac{1}{2}C_{12}^0e^1=-\omega_{02},\\
		\omega^0_{\ 3}&=C_{30}^0e^0-\frac{1}{2}C_{30}^2e^2=-\omega_{03},\ \
		\omega^1_{\ 2}=\frac{1}{2}C_{12}^0e^0-C_{12}^2e^2=\omega_{12},\\
		\omega^1_{\ 3}&=C_{31}^3e^3+C_{31}^1e^1=\omega_{13}, \ \
		\omega^2_{\ 3}=\frac{1}{2}C_{30}^2e^0+C_{32}^2e^2=\omega_{23}.
\end{align*}
Denote
\begin{align*}
		C_{10}^0&=-\sqrt{\Delta_\theta}\partial_\theta\frac{1}{\sqrt{U }},\quad C_{30}^0=\partial_r\sqrt{\frac{\Delta_r}{U }},\quad C_{12}^0=\frac{2a }{U} \sqrt{\frac{\Delta_r}{U }}\cos\theta,\\
        C_{12}^2&=\frac{1}{\sin\theta }\partial_\theta \left(\sqrt{\frac{\Delta_\theta}{U}}\sin\theta\right),\quad C_{30}^2=-\frac{2ar}{U}\sqrt{\frac{\Delta_\theta}{U}}\sin\theta, \\
        C_{31}^1&=\frac{\sqrt{\Delta_r}}{U }\partial_r\sqrt{U},\quad C_{31}^3=\frac{a^2}{U}\sqrt{\frac{\Delta_\theta}{U}}\sin\theta\cos\theta, \quad C_{32}^2=\frac{r}{U}\sqrt{\frac{\Delta_r}{U}}.
\end{align*}
Thus the spinorial connections are
\begin{align*}
		\nabla_{e_0}\Psi=&e_0\big(\Psi \big) -\frac{1}{2}\omega_{01}(e_0)e^0\cdot e^1\cdot\Psi
		-\frac{1}{2}\omega_{03}(e_0)e^0\cdot e^3\cdot\Psi\\
		&-\frac{1}{2}\omega_{12}(e_0) e^1\cdot e^2\cdot\Psi-\frac{1}{2}\omega_{23}(e_0) e^2\cdot e^3\cdot\Psi,\\
		\nabla_{e_1}\Psi=&e_1 \big(\Psi \big) -\frac{1}{2}\omega_{02}(e_1)e^0\cdot e^2\cdot\Psi
		-\frac{1}{2}\omega_{13}(e_1)e^1\cdot e^3\cdot\Psi,\\
		\nabla_{e_2}\Psi=&e_2 \big(\Psi\big)-\frac{1}{2}\omega_{01}(e_2)e^0\cdot e^1\cdot\Psi
		-\frac{1}{2}\omega_{03}(e_2)e^0\cdot e^3\cdot\Psi\\
		&-\frac{1}{2}\omega_{12}(e_2) e^1\cdot e^2\cdot\Psi-\frac{1}{2}\omega_{23}(e_2) e^2\cdot e^3\cdot\Psi,\\
		\nabla_{e_3}\Psi=&e_3 \big(\Psi \big)-\frac{1}{2}\omega_{02}(e_3)e^0\cdot e^2\cdot\Psi
		-\frac{1}{2}\omega_{13}(e_3)e^1\cdot e^3\cdot\Psi,
\end{align*}
and the Dirac operator is
\begin{align*}
D \Psi = e^\alpha \cdot \nabla_{e_\alpha}\Psi, \quad\quad \alpha =0, 1, 2, 3.
\end{align*}

Throughout the paper, we fix the following Clifford representation
\begin{equation}\label{Clifford}
	e^0\mapsto\begin{pmatrix}
		0&I\\
		I&0
	\end{pmatrix},
	\quad\quad
	e^i\mapsto\begin{pmatrix}
		0&\sigma^i\\
		-\sigma^i&0
	\end{pmatrix},
\end{equation}
where $\sigma^i$ are Pauli matrices,
\begin{equation*}
	\sigma^1=\begin{pmatrix}
		\ &1\\
		1&\
	\end{pmatrix},
	\quad\quad
	\sigma^2=\begin{pmatrix}
		\ &-i\\
		i&\
	\end{pmatrix},
	\quad\quad
	\sigma^3=\begin{pmatrix}
		1&\  \\
		\ &-1
	\end{pmatrix}.
\end{equation*}
Then, for $\alpha=0,1,2,3$, we have
\begin{eqnarray*}
	\begin{aligned}
		\nabla_{e_\alpha}\Psi=&e_\alpha \big(\Psi\big) +E_\alpha\cdot\Psi,
	\end{aligned}
\end{eqnarray*}
where
\begin{align*}
		E_0&=-\frac{1}{2}
		\begin{pmatrix}
			C_{30}^0-\frac{i C_{12}^0}{2}   &  C_{10}^0-\frac{i C_{30}^2}{2}     &  0                               &  0\\
			C_{10}^0-\frac{i C_{30}^2}{2}   & -C_{30}^0+\frac{i C_{12}^0}{2}     &  0                               &  0\\
			0                               &  0                                 & -C_{30}^0-\frac{i C_{12}^0 }{2}  & -C_{10}^0-\frac{i C_{30}^2}{2}  \\
			0                               &  0                                 & -C_{10}^0-\frac{i C_{30}^2}{2}   &  C_{30}^0+\frac{i C_{12}^0}{2}
		\end{pmatrix},\\
		E_1&=-\frac{1}{2}
		\begin{pmatrix}
			0                               &  \frac{i C_{12}^0}{2}+C_{31}^1     &  0                               &  0\\
			-\frac{i C_{12}^0}{2}-C_{31}^1  &  0                                 &  0                               &  0\\
			0                               &  0                                 &  0                               &  -\frac{i C_{12}^0}{2}+C_{31}^1  \\
			0                               &  0                                 &  \frac{i C_{12}^0}{2}-C_{31}^1   &  0
		\end{pmatrix},\\
		E_2&=-\frac{1}{2}
		\begin{pmatrix}
		    -\frac{C_{30}^2}{2}+iC_{12}^2   &  \frac{C_{12}^0}{2}-iC_{32}^2      &  0                               &  0 \\
			\frac{C_{12}^0}{2}-iC_{32}^2    &  \frac{C_{30}^2}{2}-iC_{12}^2      &  0                               &  0\\
			0                               &  0                                 & \frac{C_{30}^2}{2}+iC_{12}^2     &  -\frac{C_{12}^0}{2}-iC_{32}^2  \\
			0                               &  0                                 & -\frac{C_{12}^0}{2}-iC_{32}^2    &  -\frac{C_{30}^2}{2}-iC_{12}^2
		\end{pmatrix},\\
		E_3&=-\frac{1}{2}
		\begin{pmatrix}
			0                               &  \frac{i C_{30}^2}{2}+C_{31}^1     &  0                               &   0 \\
			-\frac{i C_{30}^2}{2}-C_{31}^3  &  0                                 &  0                               &    0\\
			0                               &  0                                 &  0                               &  -\frac{i C_{30}^2}{2}+C_{31}^3  \\
			0                               &  0                                 &  \frac{i C_{30}^2}{2}-C_{31}^3   &   0
		\end{pmatrix}.
\end{align*}

Denote 
\begin{align*}
\lambda_{\omega k}=\lambda e^{-2i\left(\omega t+(k+\frac{1}{2})\phi\right)},
\end{align*}
and
\begin{align*}
    {D_{lm}}=& (-1)^l\frac{\partial}{\partial r}+(-1)^m\frac{i}{\Delta_r}\left[\omega(r^2+a^2)+\left(k+\frac{1}{2}\right)\Xi a\right]+\frac{iQ_er}{\Delta_r},\\ 
    {L_{lm}}=&(-1)^l\frac{\partial}{\partial\theta}-\frac{(-1)^{l+m}}{\Delta_\theta}\left[a\omega\sin\theta+\frac{\Xi}{\sin\theta}
               \left(k+\frac{1}{2}\right)\right.  \\
		&\left.+(-1)^l P_e\cot\theta-(-1)^m\left(\Delta_\theta-\frac{\Xi}{2}\right)\cot\theta\right].
\end{align*}
Thus the Dirac equation \eqref{1.1} for the Majorana spinor \eqref{1.2} can be reduced to the following four differential equations \cite{ZZ}
\beq
\begin{aligned}\label{majorana eqs}
	i\lambda_{\omega k}rR_-\Theta_-        
    -\sqrt{\Delta_r}D_{11}\overline{R}_+\overline{\Theta}_+
= & a\lambda_{\omega k}\cos\theta R_-\Theta_-         
    +\sqrt{\Delta_\theta}L_{00}\overline{R}_-\overline{\Theta}_-, \\
	i\lambda_{\omega k}rR_+\Theta_+        
    +\sqrt{\Delta_r}D_{01}\overline{R}_-\overline{\Theta}_-
= & a\lambda_{\omega k}\cos\theta R_+\Theta_+         
    -\sqrt{\Delta_\theta}L_{01}\overline{R}_+\overline{\Theta}_+, \\
	i\overline{\lambda_{\omega k}}r\overline{R}_+\overline{\Theta}_+ 
    -\sqrt{\Delta_r}D_{00}{R}_-{\Theta}_-
= & -a\overline{\lambda_{\omega k}}\cos\theta \overline {R}_+\overline{\Theta}_+ 
    -\sqrt{\Delta_\theta}L_{11}{R}_+{\Theta}_+, \\
	i\overline{\lambda_{\omega k}}r\overline{R}_-\overline{\Theta}_- 
    +\sqrt{\Delta_r}D_{10}{R}_+{\Theta}_+
= & -a\overline{\lambda_{\omega k}}\cos\theta \overline {R}_-\overline{\Theta}_- 
    +\sqrt{\Delta_\theta}L_{10}{R}_-{\Theta}_-, 
\end{aligned}
\eeq
which yields
\begin{equation*}
    \Big(\beta(\theta)^2-\alpha(r)^2\Big){R}_+{\Theta}_+=\Big(\beta(\theta)^2-\alpha(r)^2\Big){R}_-{\Theta}_- =0,
\end{equation*}
where 
\begin{align*}
\alpha(r)=\frac{Q_er}{\sqrt{\Delta_r}}, \quad\quad \beta(\theta)=\frac{P_e\cot\theta}{\sqrt{\Delta_\theta}}.
\end{align*}
This gives that there are no differentiable time-periodic Majorana spinors in the Kerr-Newman and Kerr-Newman-(A)dS spacetimes if $P_e\neq0$ or $Q_e\neq0$, which is proved in \cite{ZZ}. The same result holds in the region $\Delta_r <0$ by the same argument.

Now we prove the following proposition.

\begin{prop}\label{prop2.1}
Let $\Psi$ be a differentiable solution of the Dirac equation \eqref{1.1} for the Majorana spinor \eqref{1.2} in Kerr and Kerr-(A)dS spacetimes. If $\lambda \neq 0$, then $\Psi$ must be zero.
\end{prop}
\pf Differentiating the first two equations of \eqref{majorana eqs} with respect to $\phi$ in the region $\Delta_r >0$ , we obtain
\begin{align*}
	-2i\lambda\left(k+\frac{1}{2}\right)R_-\Theta_-(a\cos\theta-ir) e^{-2i\left(\omega t+(k+\frac{1}{2})\phi\right)}=&0, \\
	-2i\lambda\left(k+\frac{1}{2}\right)R_+\Theta_+(a\cos\theta-ir)e^{-2i\left(\omega t+(k+\frac{1}{2})\phi\right)}=&0.
\end{align*}
Therefore, if $\lambda \neq 0$, then $R_+\Theta_+$ and $R_-\Theta_-$ must be zero. The result holds in the region $\Delta_r <0$ by the similar argument. 
\qed

Furthermore, we show that same results hold for $\phi$-independent or $t$-independent Majorana spinors.

\begin{prop}\label{prop2.2}
	Let ${\Psi}$ be a differentiable solution of the Dirac equation \eqref{1.1} for the Majorana spinor 
\begin{equation}\label{2.2.1}
	\Psi=S^{-1}\psi,\quad\quad  \psi={E_1}\begin{pmatrix}
		R_-(r)\Theta_-(\theta)\\
		R_+(r)\Theta_+(\theta)\\
		\overline{R}_+(r)\overline{\Theta}_+(\theta)\\
		-\overline{R}_-(r)\overline{\Theta}_-(\theta)
	\end{pmatrix},
\end{equation}
where $S$ is given by \eqref{S-matrix} and 
\begin{equation*}
%\label{E2}
	{E_1}=\text{diag}\left(e^{-i\omega t}I_{2\times2},\,\,\,\, e^{i\omega t}I_{2\times2}\right)
\end{equation*}
in Kerr and Kerr-(A)dS spacetimes. If $\lambda \neq 0$, then ${\Psi}$ must be zero.
\end{prop}
\pf
  Denote 
  \begin{align*}
  	{{D}_\pm}=&\frac{\partial}{\partial r}\pm\frac{i\omega(r^2+a^2)}{\Delta_r}, \\
  	{{L}_\pm}=&\frac{\partial}{\partial\theta}+\frac{1}{2}\left(\cot\theta+\frac{\kappa^2a^2\cos\theta\sin\theta}{\Delta_\theta}\right)\mp \frac{a\omega\sin\theta}{\Delta_\theta}.
  \end{align*}
   Thus the Dirac equation (\ref{1.1}) reduces to the following differential equations
\begin{align*}
	&i\lambda e^{-2i\omega t}rR_-\Theta_-        
	+\sqrt{\Delta_r}{D}_{+}\overline{R}_+\overline{\Theta}_+
	=a\lambda e^{-2i\omega t}\cos\theta R_-\Theta_-         
	+\sqrt{\Delta_\theta}{L}_{+}\overline{R}_-\overline{\Theta}_-, \\
	&i\lambda e^{-2i\omega t}rR_+\Theta_+        
	+\sqrt{\Delta_r}{D}_{-}\overline{R}_-\overline{\Theta}_-
	=a\lambda e^{-2i\omega t}\cos\theta R_+\Theta_+         
	-\sqrt{\Delta_\theta}{L}_{-}\overline{R}_+\overline{\Theta}_+, \\
	&i\lambda e^{2i\omega t}r\overline{R}_+\overline{\Theta}_+ 
	-\sqrt{\Delta_r}{D}_{+}{R}_-{\Theta}_-
	=-a\lambda e^{2i\omega t}\cos\theta \overline {R}_+\overline{\Theta}_+ 
	+\sqrt{\Delta_\theta}{L}_{-}{R}_+{\Theta}_+, \\
	&i\lambda e^{2i\omega t}r\overline{R}_-\overline{\Theta}_- 
	-\sqrt{\Delta_r}{D}_{-}{R}_+{\Theta}_+
	=-a\lambda e^{2i\omega t}\cos\theta \overline {R}_-\overline{\Theta}_- 
	-\sqrt{\Delta_\theta}{L}_{+}{R}_-{\Theta}_-.
\end{align*}
Differentiating the first two equations with respect to $t$ in the region $\Delta_r >0$, we obtain
\begin{align*}
	-2i\lambda\omega R_-\Theta_-(a\cos\theta-ir) e^{-2i\omega t}=&0, \\
	-2i\lambda\omega R_+\Theta_+(a\cos\theta-ir)e^{-2i\omega t}=&0.
\end{align*}
Therefore, if $\lambda \neq 0$, then $R_+\Theta_+$ and $R_-\Theta_-$ must be zero. The result holds in the region $\Delta_r <0$ by the similar argument. 
\qed

\begin{prop}\label{prop2.3}
	Let ${\Psi}$ be a differentiable solution of the Dirac equation \eqref{1.1} for the Majorana spinor 
\begin{equation}\label{2.2.2}
	\Psi=S^{-1}\psi,\quad\quad  \psi={E_2}\begin{pmatrix}
		R_-(r)\Theta_-(\theta)\\
		R_+(r)\Theta_+(\theta)\\
		\overline{R}_+(r)\overline{\Theta}_+(\theta)\\
		-\overline{R}_-(r)\overline{\Theta}_-(\theta)
	\end{pmatrix},
\end{equation}
where $S$ is given by \eqref{S-matrix} and 
\begin{equation*}
	{E_2}=\text{diag}\left(e^{-i\left(k+\frac{1}{2}\right)\phi}I_{2\times2},\,\,\,\, e^{i\left(k+\frac{1}{2}\right)\phi}I_{2\times2}\right)
\end{equation*}
in Kerr and Kerr-(A)dS spacetimes. If $\lambda \neq 0$, then ${\Psi}$ must be zero.
\end{prop}
\pf
Denote 
\begin{align*}
	{{D}_\pm}=&\frac{\partial}{\partial r}\pm\frac{ia\Xi}{\Delta_r}\left(k+\frac{1}{2}\right), \\
	{{L}_\pm}=&\frac{\partial}{\partial\theta}+\frac{1}{2}\left(\cot\theta+\frac{\kappa^2a^2\cos\theta\sin\theta}{\Delta_\theta}\right)\mp \frac{\Xi}{\Delta_\theta\sin\theta}\left(k+\frac{1}{2}\right).
\end{align*}
Thus the Dirac equation (\ref{1.1}) reduces to the following differential equations
\begin{align*}
	&\lambda e^{-2i(k+\frac{1}{2})\phi}\left(ir-a\cos\theta\right)R_-\Theta_- +\sqrt{\Delta_r}{D}_{+}\overline{R}_+\overline{\Theta}_+
	=\sqrt{\Delta_\theta}{L}_{+}\overline{R}_-\overline{\Theta}_-, \\
	&\lambda e^{-2i(k+\frac{1}{2})\phi}\left(ir-a\cos\theta\right)R_+\Theta_+ +\sqrt{\Delta_r}{D}_{-}\overline{R}_-\overline{\Theta}_-=
	-\sqrt{\Delta_\theta}{L}_{-}\overline{R}_+\overline{\Theta}_+, \\
	&\lambda e^{2i(k+\frac{1}{2})\phi}\left(ir+a\cos\theta\right)\overline{R}_+\overline{\Theta}_+ 
	-\sqrt{\Delta_r}{D}_{+}{R}_-{\Theta}_-
	=\sqrt{\Delta_\theta}{L}_{-}{R}_+{\Theta}_+, \\
	&\lambda e^{2i(k+\frac{1}{2})\phi}\left(ir+a\cos\theta\right)\overline{R}_-\overline{\Theta}_- 
	-\sqrt{\Delta_r}{D}_{-}{R}_+{\Theta}_+
	=-\sqrt{\Delta_\theta}{L}_{+}{R}_-{\Theta}_-,
\end{align*}
and the result follows by the same argument as in Proposition \ref{prop2.1}.
\qed

\begin{rmk}
	According to the above propositions, massive Majorana spinors can exist possibly in the form
	\begin{equation*}
		\Psi=S^{-1}\begin{pmatrix}
			R_-(r)\Theta_-(\theta)\\
			R_+(r)\Theta_+(\theta)\\
			\overline{R}_+(r)\overline{\Theta}_+(\theta)\\
			-\overline{R}_-(r)\overline{\Theta}_-(\theta)
		\end{pmatrix},
	\end{equation*}
	where $S$ is given by \eqref{S-matrix}. The Dirac equation \eqref{1.1} reduces to the following differential equations
	\begin{equation*}
		\begin{aligned}
			&\sqrt{\Delta_r}D\overline{R}_+\overline{\Theta}_+ +i\lambda rR_-\Theta_-=\sqrt{\Delta_\theta}L\overline{R}_-\overline{\Theta}_-+a\lambda\cos\theta R_-\Theta_-,\\
			&\sqrt{\Delta_r}D\overline{R}_-\overline{\Theta}_- +i\lambda rR_+\Theta_+=-\sqrt{\Delta_\theta}L\overline{R}_+\overline{\Theta}_+ +a\lambda\cos\theta R_+\Theta_+,\\
			&\sqrt{\Delta_r}D{R}_-{\Theta}_- -i\lambda r\overline{R}_+\overline{\Theta}_+=-\sqrt{\Delta_\theta}L{R}_+{\Theta}_+ +a\lambda\cos\theta \overline{R}_+\overline{\Theta}_+,\\
			&\sqrt{\Delta_r}D{R}_+{\Theta}_+ -i\lambda r\overline{R}_-\overline{\Theta}_-=\sqrt{\Delta_\theta}L{R}_-{\Theta}_- +a\lambda\cos\theta \overline{R}_-\overline{\Theta}_-,\\
		\end{aligned}
	\end{equation*}
	where
	\begin{equation*}
		D=\frac{\partial}{\partial r},\quad\quad L=\frac{\partial}{\partial \theta}+\frac{1}{2}\left(\cot\theta+\frac{\kappa^2a^2\cos\theta\sin\theta}{\Delta_\theta}\right).
	\end{equation*}
	We address the existence of these equations elsewhere.
\end{rmk}

\mysection{Massless Majorana spinors}

In this section, we prove the nonexistence of massless time-periodic solutions which are $L^p$ outside the event horizon for certain $p$ in the non-extreme Kerr spacetime. The Kerr metric takes the form
\begin{align*}
		ds^2=& -\left(1-\frac{2mr}{U}\right)dt^2 -\frac{4amr\sin^2\theta}{U}dtd\phi \\
		     & +\frac{U}{\Delta}dr^2+U d\theta^2 +\left(\Delta+\frac{2mr(r^2+a^2)}{U}\right)\sin^2\theta d\phi^2
\end{align*}
with
\begin{equation*}
	U=r^2+a^2\cos^2\theta,\quad\quad\Delta=r^2-2mr+a^2.
\end{equation*}
The metric is non-extreme if $m^2>a^2$. In this case, $\Delta$ has two distinct roots
\begin{equation*}
	r_c=m-\sqrt{m^2-a^2},\quad\quad r_e=m+\sqrt{m^2-a^2},
\end{equation*}
which provide Cauchy and event horizons, respectively.

Denote
\begin{align*}
{D_\pm}=&\frac{\partial}{\partial r}\pm\frac{i}{\Delta}\left[\omega(r^2+a^2)+\left(k+\frac{1}{2}\right)a\right],\\
{L_\pm}=&\frac{\partial}{\partial\theta}+\frac{\cot\theta}{2}\mp \left[a\omega\sin\theta+\frac{k+\frac{1}{2}}{\sin\theta}\right]. 
\end{align*}
For massless Majorana spinors in the Kerr spacetime, \eqref{majorana eqs} can be simplified to the following two differential equations
\begin{equation*}
	\begin{aligned}
		\sqrt{\Delta}D_-{R}_+{\Theta}_+=& L_+{R}_-{\Theta}_-,\\
		\sqrt{\Delta}D_+{R}_-{\Theta}_-=& -L_-{R}_+{\Theta}_+,
	\end{aligned}
\end{equation*}
which decouple into the following ordinary differential equations
\begin{align}
	\begin{pmatrix}
		\sqrt{\Delta}{D}_-&-\epsilon_1\\
		-\epsilon_2&\sqrt{\Delta}{D}_+
	\end{pmatrix}
	\begin{pmatrix}
		R_+\\
		R_-
	\end{pmatrix}
	= & 0, \label{2.12} \\
	\begin{pmatrix}
		-\epsilon_1&{L}_+\\
		-{L}_-&-\epsilon_2
	\end{pmatrix}
	\begin{pmatrix}
		\Theta_+\\
		\Theta_-
	\end{pmatrix}
	=& 0 \label{2.12-1}
\end{align}
for some constants $\epsilon_1$, $\epsilon_2$. Moreover, we have
\begin{align*}
	&\int_{0}^{\pi}\left(-L_-L_+\Theta_-,\Theta_{-}\right)\sin\theta d\theta=\int_{0}^{\pi}\left(L_+\Theta_-,L_+\Theta_{-}\right)\sin\theta d\theta,\\
	&\int_{0}^{\pi}\left(-L_+L_-\Theta_+,\Theta_{+}\right)\sin\theta d\theta=\int_{0}^{\pi}\left(L_-\Theta_+,L_-\Theta_{+}\right)\sin\theta d\theta,
\end{align*}
i.e.
\begin{equation}
	\begin{aligned}\label{3.5.1}
		&\overline{\epsilon_1\epsilon_2}\int_{0}^{\pi}|\Theta_{-}|^2\sin\theta d\theta=|\epsilon_1|^2\int_{0}^{\pi}|\Theta_{+}|^2\sin\theta d\theta,\\
		&\overline{\epsilon_1\epsilon_2}\int_{0}^{\pi}|\Theta_{+}|^2\sin\theta d\theta=|\epsilon_2|^2\int_{0}^{\pi}|\Theta_{-}|^2\sin\theta d\theta,
	\end{aligned}
\end{equation}
where $(\ ,\ )$ denotes the standard inner product on $\mathbb{C}^2$. Therefore $\epsilon_1\epsilon_2 $ must be real.

\begin{thm}
	In the non-extreme Kerr spacetime, for any massless time-periodic solutions \eqref{1.2} of the Dirac equation \eqref{1.1}, if $\epsilon_1\epsilon_2=0$, then \eqref{2.12} and \eqref{2.12-1} can be solved explicitly as follows.

(i) $\epsilon_1=0$ and $\epsilon_2\neq 0$,
	\begin{equation}\label{solution1}	
		\begin{aligned}
			&R_+(r)=C_1e^{i\omega r}(r-r_c)^{-i\gamma_1}(r-r_e)^{i\gamma_2},\\
			&R_-(r)=e^{-i\omega r}(r-r_c)^{i\gamma_1}(r-r_e)^{-i\gamma_2}\\
			&\quad\quad\quad\quad\times\left(C_1\epsilon_2\int e^{2i\omega r}(r-r_c)^{-2i\gamma_1-\frac{1}{2}}(r-r_e)^{2i\gamma_2-\frac{1}{2}}dr+C_2\right),\\
			&\Theta_{+}(\theta)=C_3\frac{e^{a\omega\cos\theta}}{\sqrt{\sin\theta}}\left(\tan\frac{\theta}{2}\right)^{-\left(k+\frac{1}{2}\right)},\\
			&\Theta_{-}(\theta)=0,
		\end{aligned}
	\end{equation}
	where
	\begin{equation*}
		\gamma_1=\frac{2m\omega r_c+\left(k+\frac{1}{2}\right)a}{2\sqrt{m^2-a^2}},\quad\quad\gamma_2=\frac{2m\omega r_e+\left(k+\frac{1}{2}\right)a}{2\sqrt{m^2-a^2}}.
	\end{equation*}

(ii) $\epsilon_1\neq0$ and $\epsilon_2=0$,
	\begin{equation}\label{solution2}
		\begin{aligned}
			&R_-(r)=C_4e^{-i\omega r}(r-r_c)^{i\gamma_1}(r-r_e)^{-i\gamma_2},\\
			&R_+(r)=e^{i\omega r}(r-r_c)^{-i\gamma_1}(r-r_e)^{i\gamma_2}\\
			&\quad\quad\quad\quad\times\left(C_4\epsilon_1\int e^{-2i\omega r}(r-r_c)^{2i\gamma_1-\frac{1}{2}}(r-r_e)^{-2i\gamma_2-\frac{1}{2}}dr+C_5\right),\\
			&\Theta_{+}(\theta)=0,\\
			&\Theta_{-}(\theta)=C_6\frac{e^{-a\omega\cos\theta}}{\sqrt{\sin\theta}}\left(\tan\frac{\theta}{2}\right)^{k+\frac{1}{2}}.
		\end{aligned}
	\end{equation}

(iii) $\epsilon_1=\epsilon_2=0$,
	\begin{equation}\label{solution3}
		\begin{aligned}
		&R_+(r)=C_1e^{i\omega r}(r-r_c)^{-i\gamma_1}(r-r_e)^{i\gamma_2},\\
		&R_-(r)=C_4e^{-i\omega r}(r-r_c)^{i\gamma_1}(r-r_e)^{-i\gamma_2},\\
		&\Theta_+(\theta)=C_3\frac{e^{a\omega\cos\theta}}{\sqrt{\sin\theta}}\left(\tan\frac{\theta}{2}\right)^{-\left(k+\frac{1}{2}\right)},\\
		&\Theta_-(\theta)=C_6\frac{e^{-a\omega\cos\theta}}{\sqrt{\sin\theta}}\left(\tan\frac{\theta}{2}\right)^{k+\frac{1}{2}}.
		\end{aligned}
	\end{equation}	
\end{thm}
\pf
(i) According to \eqref{3.5.1}, we have
\begin{equation*}
	\Theta_{-}(\theta)\equiv 0.
\end{equation*}
Then the equations \eqref{2.12} and \eqref{2.12-1} reduce to
\begin{equation*}
	\begin{pmatrix}
		\sqrt{\Delta}{D}_-&0\\
		-\epsilon_2&\sqrt{\Delta}{D}_+
	\end{pmatrix}
	\begin{pmatrix}
		R_+\\
		R_-
	\end{pmatrix}
	=0,\quad\quad L_-\Theta_{+}=0,
\end{equation*}
which gives \eqref{solution1}.

(ii) According to \eqref{3.5.1}, we have
\begin{equation*}
	\Theta_{+}(\theta)\equiv 0.
\end{equation*}
Then the equations \eqref{2.12} and \eqref{2.12-1} reduce to
\begin{equation*}
	\begin{pmatrix}
		\sqrt{\Delta}{D}_-&-\epsilon_1\\
		0&\sqrt{\Delta}{D}_+
	\end{pmatrix}
	\begin{pmatrix}
		R_+\\
		R_-
	\end{pmatrix}
	=0,\quad\quad L_+\Theta_{-}=0,
\end{equation*}
which gives \eqref{solution2}.

(iii) Then the equations \eqref{2.12} and \eqref{2.12-1} reduce to
\begin{equation*}
	\begin{pmatrix}
		{D}_-&0\\
		0&{D}_+
	\end{pmatrix}
	\begin{pmatrix}
		R_+\\
		R_-
	\end{pmatrix}
	=0, \quad\quad
	\begin{pmatrix}
		{L}_-&0\\
		0&L_+
	\end{pmatrix}
	\begin{pmatrix}
		\Theta_+\\
		\Theta_-
	\end{pmatrix}
	=0,
\end{equation*}
which gives \eqref{solution3}.
\qed

In the following we study the case $\epsilon_1\neq0$, $\epsilon_2\neq0$. We use the notation $a\lesssim b$ ($a\gtrsim b$) to represent $a<Cb$ ($a>Cb$) for a certain positive constant $C$. The constant is independent of the functions and parameters appearing in $a$ and $b$. 

\begin{thm}\label{thm3.1}
In the non-extreme Kerr spacetime, for any massless time-periodic solutions \eqref{1.2} of the Dirac equation \eqref{1.1} where $\epsilon_1\neq0$, $\epsilon_2\neq0$ in \eqref{2.12}, \eqref{2.12-1}, then $|R|$ is bounded at the event horizon $r=r_e$. Moreover, if the solution $\Psi$ is $L^p$ outside the event horizon for certain 
\begin{align*}
	0<p\le\frac{6}{|\epsilon_1|+|\epsilon_2| +2},
\end{align*}
$\Psi$ must be zero. In particular, any normalizable such spinor, i.e. $p=2$, must be zero if $|\epsilon_1|+|\epsilon_2|\leq 1$.
\end{thm}
\pf Let $r_1>r_e$ and denote by $M_r$ the region $r \geq r_1$ in the Kerr spacetime. We rewrite (\ref{2.12}) as
\begin{equation}
	\begin{aligned}\label{RR}
		\frac{dR_+(r)}{dr}-i\alpha_1R_+(r)-{\beta}_1R_-(r)=0,\\
		\frac{dR_-(r)}{dr}+i\alpha_1R_-(r)-{\beta}_2R_+(r)=0,
	\end{aligned}
\end{equation}
where
\begin{equation*}
	\alpha_1=\frac{1}{\Delta}\left[\omega(r^2+a^2)+\left(k+\frac{1}{2}\right)a\right],\quad \beta_1=\frac{\epsilon_1}{\sqrt{\Delta}},\quad\beta_2=\frac{\epsilon_2}{\sqrt{\Delta}}.
\end{equation*}
According to (\ref{RR}), we have
\begin{equation}\label{Rp}
	\begin{split}
		\left|\frac{d}{dr}|R|^2\right|&=\left|\beta_1 R_+\overline{R}_- +\bar{\beta}_1\overline{R}_+R_-+\beta_2 \overline{R}_+R_- +\bar{\beta}_2R_+\overline{R}_-\right|\\
		&\leq \left(|{\beta}_1|+|{\beta}_2|\right)\left(|R_+|^2+|R_-|^2\right).
	\end{split}
\end{equation}
It follows that
\begin{equation*}
	\left|\frac{d}{dr}|R|^2\right|\lesssim\left(r-r_e\right)^{-\frac{1}{2}}|R|^2,\quad\quad r\in\left(r_e,r_1\right].
\end{equation*}
Thus, outside the zero set of $R$,
\begin{equation*}
	|R(r)|\lesssim|R(s)|\exp\left(\int_{s}^{r}\left(\bar{r}-r_e\right)^{-\frac{1}{2}}\right)d\bar{r}.
\end{equation*}
Hence, we conclude that $|R|<\infty$ on $r=r_e$ and $|R|\equiv0$ on $r \in (r_e, r_1]$ if $|R|=0$ on $r=r_e$.

Moreover, according to (\ref{Rp}), we have
\begin{equation*}
	\frac{d}{dr}|R|^2\geq-\left(|{\beta}_1|+|{\beta}_2|\right)\left(|R_+|^2+|R_-|^2\right).
\end{equation*}
Therefore, there exists a sufficiently large $r_2>r_1$ such that, for $r\geq r_2$,
\begin{equation*}
	\frac{d}{dr}|R|^2\geq -\frac{|\epsilon_1|+|\epsilon_2|}{r}|R|^2,
\end{equation*}
which yields
\begin{equation*}
	|R|^2\gtrsim r^{-\left(|\epsilon_1|+|\epsilon_2|\right)}
\end{equation*}
by integration. From (\ref{3.5.1}), $\epsilon_1\epsilon_2$ is a positive real number. Let $c_1=|\epsilon_2|^2$ and $c_2=\epsilon_1\epsilon_2$. Then we have 
\begin{equation*}
	\frac{d}{dr}\left(c_1\left|R_+\right|^2-c_2\left|R_-\right|^2\right)=0
\end{equation*}
according to (\ref{RR}). This implies that
\begin{equation*}
	c_1\left|R_+\right|^2-c_2\left|R_-\right|^2=c_0
\end{equation*} 
for some constant $c_0$. If $c_0\neq 0$, without loss of generality, we can assume that $c_0>0$. Since
\begin{align*}
	|\Psi|^2&=\frac{2}{\sqrt{U\Delta}}\left(|R_+\Theta_+|^2+|R_-\Theta_-|^2\right)\\
	&=\frac{2}{\sqrt{U\Delta}}\left(\left(\frac{c_2}{c_1}|R_-|^2+\frac{c_0}{c_1}\right)|\Theta_+|^2+|R_-\Theta_-|^2\right)\\
	&\geq\frac{2}{\sqrt{U\Delta}}\frac{c_0}{c_1}|\Theta_+|^2,
\end{align*}
we have
\begin{equation*}
	\begin{split}
		\int_{M_{r_1}}|\Psi|^pd\mu&\gtrsim\int_{M_{r_1}}\left(\frac{1}{\sqrt{U\Delta}}|\Theta_+|^2\right)^{\frac{p}{2}}d\mu\\
		&\gtrsim\int_{M_{r_1}}U^{1-\frac{p}{4}}\Delta^{-\frac{1}{2}-\frac{p}{4}}\sqrt{r^2+a^2}|\Theta_+|^p\sin\theta d\phi d\theta dr\\
		&\gtrsim \int_{r_1}^{\infty}r^{2-p}dr\int_{0}^{\pi}|\Theta_+|^p\sin\theta d\theta=\infty
	\end{split}
\end{equation*}
for $0<p\le3$. If $c_0=0$, for $r\geq r_2$,
\begin{align*}
		|\Psi|^2&=\frac{2}{\sqrt{U\Delta}}\left(|R_+\Theta_+|^2+|R_-\Theta_-|^2\right)\\
		&=\frac{1}{\sqrt{U\Delta}}\left(|R_+\Theta_+|^2+|R_-\Theta_-|^2+\frac{c_2}{c_1}|R_-\Theta_+|^2+\frac{c_1}{c_2}|R_+\Theta_-|^2\right)\\
		&\gtrsim\frac{1}{\sqrt{U\Delta}}\left(|R_+|^2+|R_-|^2\right)\left(|\Theta_+|^2+|\Theta_-|^2\right)\\
		&\gtrsim \frac{1}{\sqrt{U\Delta}}r^{-\left(|\epsilon_1|+|\epsilon_2|\right)}|\Theta|^2,
\end{align*}
we have
\begin{equation*}
	\begin{split}
		\int_{M_{r_1}}|\Psi|^pd\mu&\gtrsim\int_{M_{r_2}}\left(\frac{1}{\sqrt{U\Delta}}r^{-\left(|\epsilon_1|+|\epsilon_2|\right)}|\Theta|^2\right)^{\frac{p}{2}}d\mu\\
		&\gtrsim \int_{M_{r_2}}U^{1-\frac{p}{4}}\Delta^{-\frac{1}{2}-\frac{p}{4}}r^{-\frac{p}{2}\left(|\epsilon_1|+|\epsilon_2|\right)}\sqrt{r^2+a^2}|\Theta|^p\sin\theta d\phi d\theta dr\\
		&\gtrsim \int_{r_2}^{\infty}r^{2-\left(\frac{1}{2}\left(|\epsilon_1|+|\epsilon_2|\right)+1\right)p}dr\int_{0}^{\pi}|\Theta|^p\sin\theta d\theta=\infty
	\end{split}
\end{equation*}
for $0<p\le\frac{6}{|\epsilon_1|+|\epsilon_2|+2}$. 
\qed

\mysection{Hamiltonian and self-adjointness}

In this section, we provide the Hamiltonian formulation for massless Majorana spinors. Let $u$ be a variable on $\mathbb{R}$ defined by
\begin{equation*}
	\frac{du}{dr}=\frac{r^2+a^2}{\Delta}.
\end{equation*}
The Dirac equation becomes
\begin{equation}\label{Ham}
	i\frac{\partial}{\partial t}{\psi}=H{\psi}
\end{equation}
for ${\psi}$ given by \eqref{1.2}, with the Hamiltonian
\begin{equation}\label{Hamiltonian}
	H=B\left({I}-\frac{a\sqrt{\Delta}\sin\theta}{r^2+a^2}\begin{pmatrix}
		\sigma^2&  \\
		&-\sigma^2
	\end{pmatrix}\right)\left(\hat{\mathcal{R}}+\hat{\mathcal{A}}\right),
\end{equation}
where
\begin{equation*}
	B=\left(1-\frac{a^2\Delta\sin^2\theta}{\left(r^2+a^2\right)^2}\right)^{-1},
\end{equation*}
and
\begin{align*}
	\hat{\mathcal{R}}=&\begin{pmatrix}
		\mathcal{W}_+&0&0&0\\
		0&-\mathcal{W}_-&0&0\\
		0&0&-\mathcal{W}_-&0\\
		0&0&0&\mathcal{W}_+
	\end{pmatrix},\\
	\hat{\mathcal{A}}=&\begin{pmatrix}
		0&\mathcal{M}_-&0&0\\
		\mathcal{M}_+&0&0&0\\
		0&0&0&-\mathcal{M}_-\\
		0&0&-\mathcal{M}_+&0
	\end{pmatrix},
\end{align*}
with
\begin{equation}\label{WM}
	\begin{aligned}
\mathcal{W}_\pm= & i\frac{\partial}{\partial u}\mp\frac{ia}{r^2+a^2}\frac{\partial}{\partial\phi},\\
\mathcal{M}_\pm= & \frac{\sqrt{\Delta}}{r^2+a^2}\left(i\frac{\partial}{\partial\theta}
+i\frac{\cot\theta}{2}\pm\frac{1}{\sin\theta}\frac{\partial}{\partial\phi}\right).
	\end{aligned}
\end{equation}

The following are two scalar products on spinor bundles
\begin{align}
(\Psi,\Phi)=&\overline{\Psi}\Phi,   \label{scalar product-01}\\
\langle\Psi,\Phi\rangle=&\overline{\Psi}\Phi +  \frac{a\sqrt{\Delta}\sin\theta}{r^2+a^2} \overline{\Psi} \begin{pmatrix}
		\sigma^2&  \\
		&-\sigma^2
	\end{pmatrix} \Phi. \label{scalar product-02}
\end{align}
They yield two global scalar products on spinor bundles
\begin{align}
\Big(\Psi,\Phi\Big)=&\int^\infty_{-\infty}\int^1_{-1}\int^{2\pi}_0 (\Psi,\Phi) d\phi d\cos\theta du, \label{scalar product}\\
\Big\langle\Psi,\Phi\Big\rangle=&\int^\infty_{-\infty}\int^1_{-1}\int^{2\pi}_0 \langle\Psi, \Phi \rangle d\phi d\cos\theta du.\label{2.23}
\end{align}

\begin{prop}\label{prop4.1}
	The following properties hold
	\begin{equation}\label{pro4.1.1}
		\big(e^0\Psi,\Phi\big)=\big(\Psi,e^0\Phi\big),\quad\big(e^i\Psi,\Phi\big)=-\big(\Psi,e^i\Phi\big).
	\end{equation}
	where $e^0$ and $e^i$ $(i=1, 2, 3)$ are given by \eqref{Clifford}.
\end{prop}
\pf
It follows immediately from the definition of the scalar product \eqref{scalar product-01} and the matrix representation of $e^0$, $e^i$ $(i=1, 2, 3)$.
\qed

Let $\mathcal{H}$ be the Hilbert space of spinor $\Psi$ equipped with inner product \eqref{2.23}. It can be verified that Hamiltonian \eqref{Hamiltonian} is Hermitian on $\mathcal{H}$. (But it is not the case for inner product \eqref{scalar product}.) For $u_1<u_2$, denote $\mathcal{H}_{u_1,u_2}$ the Hilbert space of spinor $\Psi$ equipped with the following inner product
\begin{equation*}
	\Big\langle\Psi,\Phi\Big\rangle_{u_1,u_2}=\int^{u_2}_{u_1}\int^1_{-1}\int^{2\pi}_0 \langle\Psi, \Phi \rangle \ d\phi d\cos\theta du,
\end{equation*}
and denote $H_{u_1,u_2}$ the restriction of Hamiltonian \eqref{Hamiltonian} on $\mathcal{H}_{u_1,u_2}$. 
\begin{prop}\label{bdy condition}
If there exist $j_1$, $j_2$ such that the following boundary conditions hold
\begin{align}
\Psi \Big| _{u=u_1} \in \mathcal{N}_{j_1,\sigma}, \quad \Psi \Big| _{u=u_2} \in \mathcal{N}_{j_2,\sigma},
\end{align}
then $H_{u_1,u_2}$ is Hermitian on $\mathcal{H}_{u_1,u_2}$, where $1\leq j_1, j_2 \leq 8$, and
	\begin{equation}\label{function space}
		\begin{aligned}
			&\mathcal{N}_{1,\pm}=\Big\{\Phi \Big| \Phi=\pm e^0\Phi\Big\}, \quad\quad \quad\,
             \mathcal{N}_{2,\pm}=\Big\{\Phi \Big| \Phi=\pm i e^3\Phi\Big\},\\
			&\mathcal{N}_{3,\pm}=\Big\{\Phi \Big| \Phi=\pm e^0e^1\Phi\Big\},\quad\quad\,\,
             \mathcal{N}_{4,\pm}=\Big\{\Phi \Big| \Phi=\pm e^0e^2\Phi\Big\},\\
            &\mathcal{N}_{5,\pm}=\Big\{\Phi \Big| \Phi=\pm ie^1e^3\Phi\Big\},\quad\quad
             \mathcal{N}_{6,\pm}=\Big\{\Phi \Big| \Phi=\pm ie^2e^3\Phi\Big\},\\
            &\mathcal{N}_{7,\pm}=\Big\{\Phi \Big| \Phi=\pm ie^0e^1e^2\Phi\Big\},\quad\,
             \mathcal{N}_{8,\pm}=\Big\{\Phi \Big| \Phi=\pm e^1e^2e^3\Phi\Big\}.
		\end{aligned}
	\end{equation}
With respect to the Clifford representation \eqref{Clifford}, it holds that
		\begin{align*}
			&\mathcal{N}_{1,\pm}=\Big\{\Phi_1=\pm \Phi_3, \Phi_2=\pm \Phi_4\Big\}, \quad\,\,\,\,
             \mathcal{N}_{2,\pm}=\Big\{\Phi_1=\pm i \Phi_3, \Phi_2=\mp i \Phi_4\Big\},\\
			&\mathcal{N}_{3,\pm}=\Big\{\Phi_1=\mp \Phi_2, \Phi_3=\pm \Phi_4\Big\},\quad\,\,\,\,
             \mathcal{N}_{4,\pm}=\Big\{\Phi_1=\pm i\Phi_2, \Phi_3=\mp i\Phi_4\Big\},\\
            &\mathcal{N}_{5,\pm}=\Big\{\Phi_1=\pm i\Phi_2, \Phi_3=\pm i\Phi_4\Big\},\quad
             \mathcal{N}_{6,\pm}=\Big\{\Phi_1=\pm \Phi_2, \Phi_3=\pm \Phi_4\Big\},\\
            &\mathcal{N}_{7,\pm}=\Big\{\Phi_1=\pm \Phi_3, \Phi_2=\mp \Phi_4\Big\},\quad\,\,\,\,
             \mathcal{N}_{8,\pm}=\Big\{\Phi_1=\mp i\Phi_3,\Phi_2=\mp i\Phi_4\Big\}.
		\end{align*}
\end{prop}
\pf 
Let 
\begin{equation*}
	\Psi=(\Psi_1,\Psi_2,\Psi_3,\Psi_4)^T,\quad\Phi=(\Phi_1,\Phi_2,\Phi_3,\Phi_4)^T.
\end{equation*}
Using \eqref{Hamiltonian} and the divergence theorem, we have
\begin{align*}
\Big\langle H\Psi,\Phi\Big\rangle &-\Big\langle \Psi,H\Phi\Big\rangle \\
	   =&\int_{u_1}^{u_2}\int_{-1}^{1}\int_{0}^{2\pi}\Big[-\left(ie^0 e^3\partial_u\Psi, \Phi\right)+\left(\Psi, ie^0 e^3\partial_u\Phi\right)\\
		&+\frac{ia}{r^2+a^2}\Big(\left(\partial_\phi \Psi, \Phi\right)+\left(\Psi, \partial_\phi \Phi\right)\Big)\\
		&+\frac{\sqrt{\Delta}}{\sin\theta(r^2+a^2)}\Big(\left(ie^0 e^2\partial_\phi \Psi, \Phi\right)-\left(\Psi, ie^0 e^2\partial_\phi \Phi\right)\Big)\\
		&+\frac{i\sqrt{\Delta}}{r^2+a^2}\Big(\left(e^0 e^1\partial_\theta\Psi, \Phi\right)+\left(\Psi, e^0 e^1\partial_\theta\Phi\right)\\
		&+\cot\theta\left(e^0 e^1\Psi, \Phi\right)\Big)\bigg]d\phi d\cos\theta du\\
		=&\int_{u_1}^{u_2}\int_{-1}^{1}\int_{0}^{2\pi}\Big[-\left(ie^0 e^3\partial_u\Psi, \Phi\right)+\left(\Psi, ie^0 e^3\partial_u\Phi\right)\\
		&+\frac{i\sqrt{\Delta}}{r^2+a^2}\Big(\left(e^0 e^1\partial_\theta\Psi, \Phi\right)+\left(\Psi, e^0 e^1\partial_\theta\Phi\right)\\
		&+\cot\theta\left(e^0 e^1\Psi, \Phi\right)\Big)\Big]d\phi d\cos\theta du\\
       =&-\frac{1}{2}\int_{-1}^{1}\int_{0}^{2\pi}\Big(\left(ie^0 e^3\Psi, \Phi\right)-\left(\Psi, ie^0 e^3\Phi\right)\Big)\Big|_{u_1}^{u_2}d\phi d\cos\theta.
\end{align*}
By Proposition \ref{prop4.1}, for $\Psi$ belonging to one of $\mathcal{N}_{j,\pm}$, it holds that
\begin{align*}
\left(ie^0 e^3\Psi, \Phi\right)=\left(\Psi, ie^0e^3\Phi\right).
\end{align*}
Thus the proposition follows. \qed

\begin{rmk}
For Chandrasekhar's separation \eqref{Dirac separation}, the boundary condition can be chosen as $\Psi(u_i)\in\mathcal{N}_{1,+}$, and, for the separation of Majorana spinors \eqref{1.2}, the boundary condition can be chosen as $\Psi(u_i)\in\mathcal{N}_{3,-}$, where $i=1,2$.
\end{rmk}

\mysection{Angular and radial equations}

In this section, we study angular and radial equations for Majorana spinors with nonzero $\epsilon _1$, $\epsilon _2$ on a radially finite interval. We provide explicit solutions of the angular equation and derive asymptotic behaviors of the radial solution at the event horizon and infinity. 

Denote $\epsilon=\epsilon_2 \neq 0$, $l^2\overline{\epsilon}=\epsilon_1 \neq 0$, where $l$ is a positive real number.

\begin{thm}\label{prop4.2}
	In the non-extreme Kerr spacetime, if ${\psi}$ is a solution of \eqref{Ham} with $\epsilon_1\neq0$, $\epsilon_2\neq0$, and satisfying the boundary condition $\Psi(u_i)\in\mathcal{N}_{3,-}$ for Majorana spinors, then $a$ must be zero and ${\psi}$ is of the form \eqref{2.2.1}. Moreover, explicit solutions of angular solution are
	\begin{equation}\label{angular solution}
		\Theta_{+}(\theta)=\Theta_0\frac{e^{-iql \theta}}{\sqrt{\sin\theta}},\quad\quad\Theta_{-}(\theta)=le^{i\theta_0}\Theta_0\frac{e^{-iql \theta}}{\sqrt{\sin\theta}}
	\end{equation}		
	for some constants $\Theta_0$.
\end{thm}
\pf The boundary condition gives
\begin{equation}\label{bdy1}
	R_-(u_1)\Theta_{-}(\theta)=R_+(u_1)\Theta_{+}(\theta).
\end{equation}
Therefore,
\begin{equation*}
	|R_-(u_1)|^2\int_{0}^{\pi}|\Theta_{-}|^2\sin\theta d\theta=|R_+(u_1)|^2\int_{0}^{\pi}|\Theta_{+}|^2\sin\theta d\theta.
\end{equation*}
According to \eqref{3.5.1}, we have 
\begin{equation*}
	l^2|R_-(u_1)|^2=|R_+(u_1)|^2.
\end{equation*}
Thus
\begin{equation*}
	\frac{R_+(u_1)}{R_-(u_1)}=\frac{\Theta_{-}(\theta)}{\Theta_{+}(\theta)}=le^{i\theta_0},
\end{equation*}
where
\begin{equation}\label{theta 0}
	\theta_0=\arg R_+(u_1)-\arg R_-(u_1).
\end{equation} 
Therefore \eqref{2.12-1} gives 
\begin{equation*}
	L_+\Theta_{+}=l\overline{\epsilon}e^{-i\theta_0}\Theta_{+},\quad L_-\Theta_{+}=-l{\epsilon}e^{i\theta_0}\Theta_{+}.
\end{equation*}
They give
	\begin{align}	      
-l\big({\epsilon}e^{i\theta_0}+\overline{\epsilon}e^{-i\theta_0}\big)\Theta_{+}=&
\big(L_--L_+\big)\Theta_{+}=2\Big(a\omega\sin\theta+\frac{k+\frac{1}{2}}{\sin\theta}\Big)\Theta_{+},\label{4.2.1}\\
-l\big({\epsilon}e^{i\theta_0}-\overline{\epsilon}e^{-i\theta_0}\big)\Theta_{+}=&
\big(L_-+L_+\big)\Theta_{+}=2\Big({\partial_\theta}+\frac{1}{2}{\cot\theta}\Big)\Theta_{+}.\label{4.2.2}
	\end{align}
It is easy to see that $\Theta_{-}$ also satisfies \eqref{4.2.1}, \eqref{4.2.2}. As the coefficient of left-hand side is constant in \eqref{4.2.1}, if  $\Theta_{+}$ or $\Theta_{-}$ is nontrivial, $a$ must be zero and ${\psi}$ must be of the form \eqref{2.2.1}. Therefore, the metric reduces to the Schwarzschild metric. Denote
\begin{equation}\label{q}
	{\epsilon}e^{i\theta_0}-\overline{\epsilon}e^{-i\theta_0}:=2q i.
\end{equation} 
Then (\ref{4.2.2}) becomes
\begin{equation*}
	\Big({\partial_\theta}+\frac{1}{2}{\cot\theta}\Big)\Theta_{+}=-iql\Theta_{+},
\end{equation*}
which yields
\begin{equation*}
	\Theta_{+}(\theta)=\Theta_0\frac{e^{-iql \theta}}{\sqrt{\sin\theta}},\quad\quad\Theta_{-}(\theta)=le^{i\theta_0}\Theta_0\frac{e^{-iql \theta}}{\sqrt{\sin\theta}}
\end{equation*}		
for some constants $\Theta_0$. 
\qed

Let 
\begin{equation}\label{XY}
	\hat{R}_-(u)=le^{i\theta_0}{R_-}(u),\quad\quad\hat{\Theta}_-(\theta)=l^{-1}e^{-i\theta_0}\Theta_{-}(\theta).
\end{equation}
Then boundary condition \eqref{bdy1} becomes
\begin{equation*}
	R_+(u_1)\Theta_{+}(\theta)=\hat{R}_-(u_1)\hat{\Theta}_{-}(\theta),
\end{equation*}
which gives
\begin{equation}\label{bdy2}
	R_+(u_1)=\hat{R}_-(u_1).
\end{equation}
Similarly, we obtain
\begin{equation}\label{bdy3}
	R_+(u_2)=\hat{R}_-(u_2).
\end{equation}

We normalize the angular solution according to 
\begin{equation*}
	\Big(\Theta,\Theta\Big)=1,
\end{equation*}
where $\Theta=(\Theta_{+}, \hat{\Theta}_{-})^T$. Since the solution of angular equation is independent of $\omega$, it needs only to consider the radial Hamiltonian equation
\begin{equation*}
	H_{r,u_1,u_2} R_{u_1,u_2}^{\omega}=\omega R_{u_1,u_2}^{\omega},
\end{equation*}
on $[u_1, u_2]$, where 
\begin{equation*}
	H_{r, u_1,u_2}=-i\sigma^3\frac{d}{du}+\frac{\sqrt{r(r-2m)}}{r^2}ql\sigma^1,
\end{equation*}
and $R_{u_1,u_2}^{\omega}=(R_{+}, \hat{R}_{-})^T$ which satisfies the boundary conditions \eqref{bdy2}, \eqref{bdy3} and the normalization condition
\begin{equation}\label{3.3}
	\Big(R^{\omega}_{u_1,u_2},R^{\omega}_{u_1,u_2}\Big)=1.
\end{equation}
The spectral decomposition of the propagator is 
\begin{equation}\label{3.5}
	e^{-itH_{r,u_1,u_2}}R_0=\sum_{\omega}e^{-i\omega t}\Big(R^{\omega}_{u_1,u_2},R_0\Big) R^{\omega}_{u_1,u_2}.
\end{equation}

In the following we study the asymptotic behaviors of the radial solution when $u \rightarrow \mp \infty$. Denote $l^{*}=l{\epsilon}e^{i\theta_0}$. The radial equation \eqref{2.12} can be written as
\begin{equation}\label{3.7}
	\left[\frac{d}{du}-i\omega\begin{pmatrix}
		1&0\\
		0&-1
	\end{pmatrix}
	\right]\begin{pmatrix}
		R_+\\
		\hat{R}_-
	\end{pmatrix}=\frac{\sqrt{r(r-2m)}}{r^2}\begin{pmatrix}
		0&\bar{l}^{*}\\
		l^{*}&0
	\end{pmatrix}\begin{pmatrix}
	R_+\\
	\hat{R}_-
	\end{pmatrix}.
\end{equation}

The next four lemmas are essentially the same as those in \cite{FK}. As coefficients in \eqref{3.7} are slightly different from the equation in \cite{FK}, we present the proof here for convenience. Denote $g:=(g^+,g^-)^T$, and let
\begin{equation*}
	\begin{pmatrix}
		R_+(u)\\
		\hat{R}_-(u)
	\end{pmatrix}=\begin{pmatrix}
		e^{i\omega u}g^+(u)\\
		e^{-i\omega u}g^-(u)
	\end{pmatrix}.
\end{equation*}
By \eqref{3.7}, $g$ satisfies
\begin{equation}\label{3.10}
	\frac{d}{du}g=
		\frac{\sqrt{r(r-2m)}}{r^2}
		\begin{pmatrix}
			0&\bar{l}^{*}e^{-2i\omega u}\\
			l^{*}e^{2i\omega u}&0
		\end{pmatrix}g.
\end{equation}
Note that the coefficient matrix of right-hand side of \eqref{3.10} vanishes as $u\rightarrow{-\infty}$. 

\begin{lem}\label{lem3.1}
Let $R$ be a solution of \eqref{3.7} with $\epsilon_1\neq0$, $\epsilon_2\neq0$. Then there exist nontrivial $g_0$ and positive constant $d$ such that
\begin{equation*}
	\begin{pmatrix}
		R_+(u)\\
		\hat{R}_-(u)
	\end{pmatrix}=\begin{pmatrix}
		e^{i\omega u}g^+_0\\
		e^{-i\omega u}g^-_0
	\end{pmatrix}
	+E_0(u), \quad |E_0|\lesssim e^{du}.
\end{equation*}
\end{lem}
\pf
By \eqref{3.10}, we have
\begin{equation}\label{3.13}
	\Big|\frac{d}{du}g\Big|\lesssim e^{du}|g|
\end{equation}
for all $u\le u_2$. Integrating it, we obtain
\begin{equation*}
	\Big|\ln|g|\Big|^{u_2}_{u}\Big|\lesssim \frac{1}{d} e^{du}\Big|^{u_2}_u.
\end{equation*}
Thus, $|g|$ is bounded away from zero, and
\begin{equation}\label{3.15}
	\Big|\frac{d}{du}g\Big|\lesssim e^{du}.
\end{equation}
Integrating \eqref{3.15}, we obtain
\begin{equation*}
	\big|E_0(u)\big|=\big|g(u)-g_0\big|=\Big|\int_{-\infty}^{u}\frac{d}{d\bar{u}}g \ d\bar{u}\Big|\le\int_{-\infty}^{u}\big|\frac{d}{d\bar{u}}g\Big| \bar{u}\lesssim e^{du}.
\end{equation*}
\qed

Lemma \ref{lem3.1} shows that $R(u)$ does not vanish as $u\rightarrow{-\infty}$. Thus, the norm of $R^{\omega}_{-\infty,u_2}$ is not finite. We normalize it according to
\begin{equation}
	\lim\limits_{u\rightarrow{-\infty}}\big|R^{\omega}_{-\infty,u_2}\big|=1. \label{normalization-2}
\end{equation}

\begin{lem}\label{lem3.2}
Let $R$ be a solution of \eqref{3.7} with $\epsilon_1\neq0$, $\epsilon_2\neq0$. For fixed $u_2$, asymptotically as $u_1\rightarrow{-\infty}$, it holds that
\begin{align}
	R^{\omega}_{u_1,u_2}&=f(u_1)R^{\omega}_{-\infty,u_2}\Big|_{[u_1,u_2]}, \label{3.19}\\
	|f(u_1)|^{-2} &=(u_2-u_1)+\mathcal{O}(1). \label{3.20}
\end{align}
\end{lem}
\pf
Since $R^{\omega}_{u_1,u_2}$ and $R^{\omega}_{-\infty,u_2}$ are solutions of the \eqref{3.7} with the same boundary condition \eqref{bdy3} at $u_2$, we have
\begin{align*}
R^{\omega}_{u_1,u_2}=f(u_1)R^{\omega}_{-\infty,u_2}\Big|_{[u_1,u_2]}
\end{align*}
with a normalization factor $f$. According to \eqref{3.3}, we have
\begin{equation*}
    \int_{u_1}^{u_2}\big|R^{\omega}_{u_1,u_2}\big|^2du=\big|f(u_1) \big|^2 \int_{u_1}^{u_2}\big|R^{\omega}_{-\infty,u_2}\big|^2du=1.
\end{equation*}
Therefore
\begin{equation*}
	\big|f(u_1) \big|^{-2}=\int_{u_1}^{u_2}\Big(1+\overline{R}E_0+\overline{E}_0R+|E_0|^2\Big)du.
\end{equation*}
Since $R$ is bounded and $E_0$ has exponential decay as $u \rightarrow -\infty$, \eqref{3.20} follows.
~\hfill\qed

 \begin{lem}\label{lem3.3}
Let $R$ be a solution of \eqref{3.7} with $\epsilon_1\neq0$, $\epsilon_2\neq0$. Denote 
\begin{align*}
\Delta\omega=\min\Big\{{\omega}^\prime-\omega \big|{\omega}^\prime>\omega\Big\}. 
\end{align*}
For fixed $u_2$ locally uniformly in $\omega$, we have
\begin{equation}
	\Delta \omega=\frac{\pi}{u_2-u_1}+\mathcal{O}\Big((u_2-u_1)^{-2}\Big) \label{lemma2-3}
\end{equation}
asymptotically as $u \rightarrow -\infty$.
\end{lem}
\pf
By \eqref{3.7}, we have
\begin{equation*}
	\frac{d}{du}\Big(\big|R_+\big|^2-\big|\hat{R}_- \big|^2\Big)=0.
\end{equation*}
Thus, for all $u\le u_2$,
\begin{align*}
\big|R_+ \big|^2=\big|\hat{R}_- \big|^2.
\end{align*}
So we can assume
\begin{equation}\label{U1}
	\arg R_+(u_1)=\arg \hat{R}_-(u_1).
\end{equation}
Differentiating \eqref{3.10} with respect to $\omega$, we obtain
\begin{equation*}
	\Big|\frac{d}{du}\partial_\omega g\Big|\lesssim e^{du} \big|\partial_\omega g \big|+ e^{du} \big|g \big|.
\end{equation*}
Since $|g|$ is bounded, we have
\begin{equation*}
	\Big|\ln\big( \big|\partial_\omega g \big|+c\big)\Big|^{u_2}_u\Big|\lesssim \frac{1}{d} e^{du}\Big|^{u_2}_u
\end{equation*}
for some positive constant $c$. Thus $|\partial_\omega g|$ is bounded.

Denote the phase function
\begin{equation}\label{3.30}
	h(u)=\arg g_+(u)-\arg g_-(u)+2\omega u_2.
\end{equation}
Differentiating \eqref{3.30} with respect to $u$ and $\omega$, we conclude that
\begin{equation*}
	\Big|\frac{d}{du}\partial_\omega h\Big|\lesssim e^{du}.
\end{equation*}
Since $h(u_2)=0$, it follows that $|\partial_\omega h(u)|$ is bounded. 

The boundary condition \eqref{U1} holds if and only if
\begin{equation*}
	H_0:=2\omega(u_1-u_2)+h=0\,\, (\text{mod} \ 2\pi).
\end{equation*} 
With the above estimate, we obtain
\begin{equation*}
	\Big|H_0(\omega_{2})-H_0(\omega_{1})-2(\omega_{2}-\omega_1)(u_1-u_2)\Big|\leq\int_{\omega_{1}}^{\omega_{2}}\big|\partial_\omega h \big|d\omega \leq C(\omega_{2}-\omega_{1}).
\end{equation*}
As the estimate holds as $u_1 \rightarrow -\infty$, it must hold that
\begin{align*}
H_0(\omega_{2})-H_0(\omega_{1})=-2\pi.
\end{align*}
Therefore,
\begin{align*}
\Big|2\pi-2\Delta\omega(u_2-u_1)\Big|\leq C\Delta\omega,
\end{align*}
for positive constant $C$. This gives that, as $u_1 \rightarrow -\infty$,
\begin{equation*}
	\frac{2\pi}{2(u_2-u_1)+C}\leq \Delta\omega\leq\frac{2\pi}{2(u_2-u_1)-C}.
\end{equation*}
Thus, \eqref{lemma2-3} holds.\qed

Now we study asymptotic behavior of the radial solution as $u\rightarrow{+\infty}$. The radial equation takes the form
\begin{equation}\label{3.35}
	\frac{d}{du}\begin{pmatrix}
		R_+(u)\\
		\hat{R}_-(u)
	\end{pmatrix}=T(u)\begin{pmatrix}
	R_+(u)\\
	\hat{R}_-(u)
	\end{pmatrix},
\end{equation}
and, as $u\rightarrow{+\infty}$,
\begin{equation*}
T(u) \longrightarrow T_\infty:=\begin{pmatrix}
		i\omega&  \\
		 &-i\omega
	\end{pmatrix}.
\end{equation*}
Since eigenvalues of $T_\infty$ are $\mu=\pm i\omega$, which are pure imaginary, solutions are oscillatory. Denote $R_{1,2}^{\omega}$ the solutions which are asymptotic to
\begin{equation*}
	g^{\omega}_{0,1}=\begin{pmatrix}
		1\\
		0
	\end{pmatrix},
	\quad\quad\quad
	g^{\omega}_{0,2}=\begin{pmatrix}
		0\\
		1
	\end{pmatrix}.
\end{equation*}
respectively as $u\rightarrow{-\infty}$.

\begin{lem}\label{lem3.4}
Let $R$ be a solution of \eqref{3.7} with $\epsilon_1\neq0$, $\epsilon_2\neq0$. Then there exists nontrivial $g_\infty$ such that
	\begin{equation*}
		\begin{pmatrix}
			R_+(u)\\
			\hat{R}_-(u)
		\end{pmatrix}=\begin{pmatrix}
			e^{i\omega u}g^+_\infty\\
			e^{-i\omega u}g^-_\infty
		\end{pmatrix}
		+E_\infty(u), \quad \big|E_\infty \big|\lesssim\frac{1}{u}.
	\end{equation*}
\end{lem}
\pf For $T(u)$ in \eqref{3.35}, there exists matrix $D$ such that
\begin{equation*}
	D^{-1}TD=i\delta(u) \sigma^3
\end{equation*}
for some suitable function $\delta(u)$. As $u\rightarrow{+\infty}$,
\begin{equation*}
	T(u)=T_\infty+\frac{1}{u}T_1+\mathcal{O}\left(u^{-2}\right),
\end{equation*}
we can choose $D$ such that $|D(u)|$ is bounded and
\begin{equation}\label{3.41}
	 \big|D^\prime(u)\big|\lesssim\frac{1}{u^2}.
\end{equation}
Therefore, $D^{-1}R$ satisfies
\begin{equation*}
	\frac{d}{du}\big (D^{-1}R\big) = \big[i\delta\sigma^3-D^{-1}D^\prime\big]\big(D^{-1}R\big).
\end{equation*}
Let
\begin{equation*}
	R=D\begin{pmatrix}
		e^{iw(u)}g^+(u)\\
		e^{-iw(u)}g^-(u)
	\end{pmatrix},
	\quad
	w^\prime(u)=\delta(u)
\end{equation*}
and substituting \eqref{3.41}, we obtain
\begin{equation*}
	\Big|\frac{d}{du} g\Big|\lesssim \frac{1}{u^2}\big|g\big|.
\end{equation*}
A direct calculation shows that
\begin{align*}
	w(u)=&\omega u,\\
	D(u)=&I+\mathcal{O}(u^{-1}).
\end{align*}
The term of order $\mathcal{O}(u^{-1})$ can be absorbed into $E_\infty$. Using similar discussion as in Lemma \ref{lem3.1}, we can prove that $|g(u)|$ is bounded away from zero, and
\begin{equation}\label{3.46}
	\Big|\frac{d}{du} g\Big|\lesssim\frac{1}{u^2}.
\end{equation}
Thus, $g$ has a nonzero limit $g_\infty$ as $u\rightarrow \infty$. Integrating \eqref{3.46}, the estimate of $E_{\infty}$ follows.
\qed

\mysection{Integral representation and decay of the Probability}\ls

In this section, we adopt the argument in \cite{FK} to prove the integral representation of $\exp(-itH_r)$ and derive the decay of the probability for massless Majorana spinors.
\begin{thm}\label{thm3.6}
	Let $R$ be a solution of \eqref{3.7} with $\epsilon_1\neq0$, $\epsilon_2\neq0$. For Cauchy data $R_0\in C_0^\infty\left(\mathbb{R}\right)$, 
	\begin{equation*}
e^{-itH_{r}}R_0=\frac{1}{\pi} \int_{-\infty}^{\infty}e^{-i\omega t} \sum_{a,b=1}^{2} t_{ab}^{\omega}\Big(R_b^{\omega},R_0 \Big) R_a^{\omega}d\omega,
	\end{equation*}
	where
	\begin{equation}\label{3.48}
		{t_{ab}^\omega}=\frac{1}{2\pi}\int_{0}^{2\pi}\frac{{t_a}\bar{t}_b}{|t_1|^2+|t_2|^2}d\alpha
	\end{equation}
	with
	\begin{equation*}
		t_1(\alpha)=g^{+}_{\infty 2}e^{i\alpha}-g^{-}_{\infty 2} e^{-i\alpha},\quad t_2(\alpha)=-g^{+}_{\infty 1}e^{i\alpha}+g^{-}_{\infty 1} e^{-i\alpha}.
	\end{equation*} 
\end{thm}
\pf By Lemma \ref{lem3.2}, \eqref{3.5} becomes
\begin{equation}\label{6.1.2}
	e^{-itH_{r,u_1,u_2}}R_0=\frac{1}{(u_2-u_1)+\mathcal{O}(1)}\sum_{\omega}e^{-i\omega t}
\Big( R_{-\infty,u_2}^{\omega},R_0\Big) R_{-\infty,u_2}^{\omega}.
\end{equation}
By Lemma \ref{lem3.3}, the right-hand side of \eqref{6.1.2} converges to an integral as $u_1\rightarrow{-\infty}$
\begin{equation*}
e^{-itH_{r,-\infty,u_2}}R_0=\frac{1}{\pi}\int_{-\infty}^{\infty}e^{-i\omega t}\Big( R_{-\infty,u_2}^{\omega},R_0\Big)
R_{-\infty,u_2}^{\omega}d\omega.
\end{equation*}
At $t=0$,
\begin{equation*}
	R_0=\frac{1}{\pi}\int_{-\infty}^{\infty}\Big( R_{-\infty,u_2}^{\omega},R_0\Big) R_{-\infty,u_2}^{\omega}d\omega.
\end{equation*}
Furthermore, for positive constant $\tau$,
\begin{equation}\label{3.49}
	R_0=\frac{1}{\pi}\int^\infty_{-\infty}\Big( R_{-\infty,u_2+\tau}^{\omega},R_0\Big) R^{\omega}_{-\infty,u_2+\tau}d\omega.
\end{equation}
Taking the average over $\tau$ in the interval $[0,T]$ with $T > 0$, we obtain
\begin{equation}\label{3.50}
R_0=\frac{1}{\pi}\int^\infty _{-\infty}\bigg[\frac{1}{T}\int^T_0 \Big( R_{-\infty,u_2+\tau}^{\omega},R_0\Big) R^{\omega}_{-\infty,u_2+\tau}d\tau\bigg]d\omega.
\end{equation}
Let
\begin{equation}\label{IT}
	I(T)=\frac{1}{T}\int^T_0\Big( R_{-\infty,u_2+\tau}^{\omega},R_0\Big) R^{\omega}_{-\infty,u_2+\tau}d\tau.
\end{equation}
Since 
\begin{equation*}
	R^{\omega}_{-\infty,u_2+\tau}=\sum\limits^2_{a=1}c_a(\tau)R^{\omega}_{a}
\end{equation*}
and the boundary condition \eqref{bdy3} is satisfied at $u_2+\tau$, for sufficiently large $u_2$, \eqref{IT} becomes
\begin{equation}\label{3.51}
	\sum^2_{a,b=1}t_{ab}\Big( R^{\omega}_b,R_0\Big) R^{\omega}_a,\quad t_{ab}=\frac{1}{T}\int^T_0 c_a(\tau)\overline{c_b(\tau)}d\tau
\end{equation}
and the boundary condition implies
\begin{equation}\label{3.52}
	\begin{split}
		& c_1(\tau)\Big(g^+_{\infty 1} e^{iw(u_2+\tau)}-g^-_{\infty 1} e^{-iw(u_2+\tau)}\Big)\\
		&+c_2(\tau)\Big(g^+_{\infty 2} e^{iw(u_2+\tau)}-g^-_{\infty 2} e^{-iw(u_2+\tau)}\Big)+\mathcal{O}\left(\tau^{-1}\right)=0.
	\end{split}
\end{equation}
By normalization, we have
\begin{equation*}
	|c_1|^2+|c_2|^2=1.
\end{equation*}
Therefore, the solution of \eqref{3.52} is
\begin{equation}\label{3.54}
	\begin{pmatrix}
		c_1\\
		c_2
	\end{pmatrix}=
	\frac{1}{Z}
	\begin{pmatrix}
		g^+_{\infty 2} e^{iw(u_2+\tau)}-g^-_{\infty 2} e^{-iw(u_2+\tau)}\\
		-g^+_{\infty 1} e^{iw(u_2+\tau)}+g^-_{\infty 1} e^{-iw(u_2+\tau)}
	\end{pmatrix}+\mathcal{O}\left(\tau^{-1}\right),
\end{equation}
where $Z$ is a normalization parameter. Considering $c_1$, $c_2$ as functions of $w$ and substituting \eqref{3.54} into \eqref{3.51}, it follows that
\begin{equation*}
	t_{ab}(T)=\frac{1}{T}\int^{w(T)}_{w(0)}c_a(w)\overline{c_b(w)} \ \frac{dw}{|w^\prime|}.
\end{equation*}
Hence, by Lemma \ref{lem3.4}, $t_{ab}(T)$ converges to \eqref{3.48} as $T\rightarrow\infty$ and $I(T)$ converges pointwisely. Next, we show that
\begin{equation*}
	|I(T)|\lesssim G(\omega),
\end{equation*}
where $G(\omega)$ is an integrable function. According to
\begin{equation*}
	R^{\omega}_{-\infty,u_2+\tau}=\begin{pmatrix}
		R_+(u)\\
		\hat{R}_-(u)
	\end{pmatrix}=\begin{pmatrix}
		e^{i\omega u}g^+(u)\\
		e^{-i\omega u}g^-(u)
	\end{pmatrix},
\end{equation*}
we have 
\begin{equation*}
	\big|R^{\omega}_{-\infty,u_2+\tau}(u)\big|=\big|g(u)\big|.
\end{equation*}
By Lemma \ref{lem3.1}, there exists $u_0<u_2$ such that, for $u\leq u_0$, $|g(u)|$ is bounded. According to \eqref{3.10},
\begin{equation*}
	\Big|\frac{d}{du} \big|g \big|^2 \Big|\lesssim \big|g \big|^2
\end{equation*}
for $u\in[u_0,u_2+\tau]$. Therefore, for all $u\in(-\infty,u_2+\tau]$, we have
\begin{equation*}
	\big|R^{\omega}_{-\infty,u_2+\tau}(u)\big|=\big|g(u) \big|\leq L_1
\end{equation*}
for some positive constant $L_1$. Choosing $n_0$ sufficiently large such that $\text{supp} \ R_0\subset[-n_0,n_0]$, we obtain
\begin{equation*}
	\Big( R^{\omega}_{-\infty,u_2+\tau},R_0\Big) =\int_{-n_0}^{n_0}\Big(e^{-i\omega u}\overline{g^+}R_0^+ +e^{i\omega u}\overline{g^-}R_0^-\Big)du,
\end{equation*}
where $R_0=\left(R_0^+,R_0^-\right)^T$. Since $R_0$ is compactly supported, its Fourier transform decays rapidly
\begin{equation*}
	\Big|\Big( R^{\omega}_{-\infty,u_2+\tau},R_0\Big) \Big|\lesssim\frac{C_N}{(|\omega|+1)^N}
\end{equation*}
for any $N>1$. Defining
\begin{equation*}
	G(\omega)=\frac{C_N}{(|\omega|+1)^N},
\end{equation*}
Lebesgue's dominated convergence Theorem implies
\begin{align*}
%\label{6.1.1}
		\lim_{T\rightarrow\infty}\frac{1}{\pi} & \int^\infty_{-\infty}
\bigg[\frac{1}{T}\int^T_0\Big( R^{\omega}_{-\infty,u_2+\tau},R_0\Big) R^{\omega}_{-\infty,u_2+\tau}d\tau\bigg]d\omega\\
		=\frac{1}{\pi} &\int_{-\infty}^{\infty} \sum_{a,b=1}^{2} t_{ab}^{\omega}\Big( R_b^{\omega},R_0\Big) R_a ^{\omega}d\omega.
\end{align*}
The proof of theorem is completed by applying $\exp(-it H_r)$ on both sides. ~\hfill\qed

Now, we show the decay of the probability. 
\begin{thm}
Consider Cauchy problem for the Dirac equation in the Schwarzschild spacetime 
\begin{equation*}
	D\Psi=0,\quad\Psi(0,x)=\Psi_0, 
\end{equation*}
with the initial data $\Psi_0 \in L^2\big((2m,\infty)\times S^2, dvol\big)$, where $dvol$ is the volume form on the hypersurface $t=\text{const}$. For any $\delta>0$, $\mu>2m+\delta$, the probability of massless Majorana spinors with $\epsilon_1\neq 0$, $\epsilon_2\neq 0$, $|\epsilon_1|+|\epsilon_2|>1$ to be inside the annulus 
\begin{align*}
K_{\delta,\mu}=\Big\{2m+\delta\le r\le \mu \Big\}
\end{align*}
tends to zero as time tends to infinity, i.e.,
	\begin{equation*}
		\lim\limits_{t\rightarrow\infty}\int_{K_{\delta,\mu}}|\Psi|^2dvol=0.
	\end{equation*}
\end{thm}
\pf
For any $\varepsilon>0$, there exists ${R}_I\in C_0^\infty(2m,\infty)$ such that
\begin{equation*}
	\Big\Vert{R}_{I}-{R}_0\Big\Vert<\varepsilon.
\end{equation*}
According to Theorem \ref{thm3.6}, we have
\begin{equation}\label{4.3}
    {R}_{I}=\frac{1}{\pi}\int^\infty_{-\infty} t^{\omega}_{ab}\Big( R^{\omega}_b,{R}_{I}\Big) R^{\omega}_a d\omega.
\end{equation}
Since the integrand in \eqref{4.3} is bounded, it is in $L^1$ as a function of $\omega$. Therefore, the corresponding Fourier transform is in $L^\infty$ with respect to $t$, and it tends to zero as $t\rightarrow \infty$ by Riemann-Lebesgue Lemma \cite{RS}. Thus, 
\begin{equation*}
	\lim\limits_{t\rightarrow \infty}e^{-it H_r}{R}_{I}=0,
\end{equation*}
where
\begin{equation*}
    e^{-itH_r}{R}_{I}=\frac{1}{\pi}\int^\infty_{-\infty} e^{-i\omega t}t^{\omega}_{ab}\Big( R^{\omega}_b,{R}_{I}\Big) R^{\omega}_ad\omega.
\end{equation*}
On annulus $K_{\delta,\mu}$, 
\begin{equation*}
		e^{-itH_r}{R}_0=e^{-itH_r}\big({R}_0-{R}_I\big)+e^{-itH_r}{R}_I,
\end{equation*}
therefore, there is a constant $c\geq 1$ depending only on $\delta$ and $\mu$ such that
\begin{equation*}
\Big( e^{-itH_r}R_0, e^{-itH_r}R_0 \Big) _{K_{\delta,\mu}}
	\le\Big( e^{-itH_r}R_I,e^{-itH_r}R_I \Big) _{K_{\delta,\mu}}+c\varepsilon^2+2c\varepsilon \Big\Vert R_{I}\Big\Vert.
\end{equation*}
This implies that the integral converges to zero by applying the Lebesgue's dominated convergence Theorem. 
\qed

\bigskip

\footnotesize {

\noindent {\bf Acknowledgement} The work is supported by the National Natural Science Foundation of China 12326602.

}


\begin{thebibliography}{99}

\bibitem{AA} Alavi, S. A., Abbasnezhad, A. Can gravity distinguish between Dirac and Majorana neutrinos? Gravitation Cosmology, 2016, 22, 288–298.
\bibitem{BS06} Batic, D., Schmid, H. The Dirac propagator in the Kerr-Newman metric. Progress of Theoretical Physics, 2006, 116(3), 517-544.
\bibitem{BS07} Batic, D., Schmid, H. The Dirac propagator in the extreme Kerr metric. Journal of Physics A: Mathematical and Theoretical, 2007, 40, 13443–13451.
\bibitem{B07} Batic, D. Scattering for massive Dirac fields on the Kerr metric. Journal of Mathematical Physics, 2008, 48, 022502.
\bibitem{BC09} Belgiorno, F., Cacciatori, S. L. The absence of normalizable time-periodic solutions for the Dirac equation in the Kerr-Newman-dS black hole background. Journal of Physics A: Mathematical and Theoretical, 2009, 42(13): 135207.
\bibitem{BC10} Belgiorno, F., Cacciatori, S. L. The Dirac equation in Kerr-Newman-AdS black hole background. Journal of Mathematical Physics, 2010, 51: 033517.
\bibitem{SC} Chandrasekhar, S. The solution of Dirac’s equation in Kerr geometry. Proceedings of Royal Society of London Series A-Mathematical and Physics Sciences, 1976, 349: 571–575.
\bibitem{FWZ24} Fan, M., Wang, Y., Zhang, X. Nonexistence of time-periodic solutions of the Dirac equation in Kerr-Newman-(A)dS spacetime. arXiv:2404.13255v1 [gr-qc] 20 Apr 2024.
\bibitem{FJ} Finster, F., Kamran, N., Smoller, J., Yau, S. T. Non-existence of time-periodic solutions of the Dirac equation in an axisymmetric black hole geometry. Communications on Pure and Applied Mathematics, 2000, 53: 902–929.
\bibitem{FK} Finster, F., Kamran, N., Smoller, J., Yau, S. T. The long-time dynamics of Dirac particles in the Kerr-Newman black hole geometry. Advances in Theoretical and Mathematical Physics, 2000, 7(1): 25-52.
\bibitem{FF} Finster, F., Kamran, N., Smoller, J., Yau, S. T. Decay Rates and Probability Estimates for Massive Dirac Particles in the Kerr–Newman Black Hole Geometry. Communications in Mathematical Physics, 2002, 230(2): 201-244.
\bibitem{FC} Finster, F., Krpoun, C. An integral representation for the Dirac propagator in the Reissner-Nordström geometry in Eddington–Finkelstein coordinates. Letters in Mathematical Physics, 2025, 115, 59.
\bibitem{GT} Garavaglia, T. Dirac- and Majorana-neutrino-mass effects in neutrino-electron elastic scattering. Physical Review D, 1984, 29: 3(3): 387-392.
\bibitem{HM} Hsieh, C. L., Memari, V., Halilsoy, M. The Majorana Hopping and Constraints in the Anti-de Sitter Spacetime. 2024, arXiv: 2403.13810v2.
\bibitem{EM} Majorana, E. Teoria simmetrica dell’elettrone e del positrone. Nuovo Cimento C-Colloquia and Communications in Physics, 1937, 14, 171.
\bibitem{MS} Murthy, M. V. N., Sahoo, D., Kim, C. S. Inferring the nature of active neutrinos: Dirac or Majorana? Physical review D, 2022, 105(11 Pt.A): 113006.
\bibitem{NJ} Nieves, J. F., Pal, P. B. Minimal rephasing-invariant CP-violating parameters with Dirac and Majorana fermions. Physical Review D, 1987, 36(1): 315.
\bibitem{PD} Page, D. Dirac equation around a charged, rotating black hole. Physical Review D, 1976, 14(6): 1509–1510.
\bibitem{PU} Prokopec, T., Unnithan, V. H. Majorana propagator on de Sitter space. The European Physical Journal C, 2022, 82(11).
\bibitem{RS} Reed, M., Simon, B. Methods of Modern Mathematical Physics II: Fourier Analysis, Self-adjointness. New York: Academic Press, 1975: 10.
\bibitem{RW} Rodejohann, W. Neutrino-less Double Beta Decay and Particle Physics. International Journal of Modern Physics E, 2012, 20(9).
\bibitem{WZ18} Wang, Y., Zhang, X. Nonexistence of time-periodic solutions of the Dirac equation in non-extreme Kerr-Newman-AdS spacetime. Science China Mathematics, 2018, 61(1): 73-82.
\bibitem{ZZ} Zhang, H., Zhang, X. Nonexistence of Majorana fermions in Kerr-Newman type spacetimes with nontrivial charge. Chinese Physics C, 2024, 48(11): 115104.
\bibitem{ZZ1} Zhang, H., Zhang, X. In preparation.
\end{thebibliography}
\end{document}